\documentclass[aps,prl,twocolumn,superscriptaddress]{revtex4-2}
\usepackage{graphicx}
\usepackage{xcolor}
\usepackage{hyperref}
\usepackage[normalem]{ulem}

\begin{document}
\title{Electron acceleration by laser plasma wedge interaction}
\date{\today}

\author{S. Marini} 
\affiliation{LSI, CEA/DRF/IRAMIS, CNRS, \'Ecole Polytechnique, Institut Polytechnique de Paris, F-91128 Palaiseau, France.}
\affiliation{LULI, Sorbonne Universit\'e, CNRS, CEA, \'Ecole Polytechnique, Institut Polytechnique de Paris, F-75252 Paris, France.}

\author{P. S. Kleij} 
\affiliation{LSI, CEA/DRF/IRAMIS, CNRS, \'Ecole Polytechnique, Institut Polytechnique de Paris, F-91128 Palaiseau, France.}

\author{M. Grech} 
\affiliation{LULI, CNRS, CEA, Sorbonne Universit\'e, \'Ecole Polytechnique, Institut Polytechnique de Paris, F-91120 Palaiseau, France.}

\author{M. Raynaud} 
\affiliation{LSI, CEA/DRF/IRAMIS, CNRS, \'Ecole Polytechnique, Institut Polytechnique de Paris, F-91128 Palaiseau, France.}

\author{C. Riconda \footnote{caterina.riconda@upmc.fr}}
\affiliation{LULI, Sorbonne Universit\'e, CNRS, CEA, \'Ecole Polytechnique, Institut Polytechnique de Paris, F-75252 Paris, France.}

\begin{abstract}
A new electron acceleration mechanism is identified that develops when a relativistically intense laser irradiates the wedge of an over-dense plasma. This induces a diffracted electromagnetic wave with a significant longitudinal electric field that accelerates electrons from the plasma over long distances to relativistic energies. Well collimated, highly-charged (nC) electron bunches with energies up to 100's MeV are obtained using a laser beam with $I \lambda_0^2 =3.5\times 10^{19}\,{\rm W \mu m^2/cm^2}$. Multi-dimensional particle-in-cell simulations, supported by a simple analytical model, confirm the efficiency and robustness of the proposed acceleration scheme.
\end{abstract}

\maketitle
The possibility of developing new compact energetic particle and radiation sources via several mechanisms involving the interaction of an ultra intense laser and plasmas has gained importance in the last decades, given the numerous applications that range from image generation \citep{joy01} to proton-therapy \citep{bulanov14}, passing through space propulsion \citep{levchenko20}. In order to promote the particle acceleration, various designs were proposed and studied in detail, either involving the broad category of Laser Wakefield Acceleration (LWFA) \citep{esarey95}, or 
the interaction of a laser with an overdense plasma \citep{raynaud07,riconda15,macchi18,cantono18,raynaud20,marini21,marini21b,lundh11,grandvaux20,knyazev21,quesnel98,cantono18,marini17,marini21,esarey09,hartemann95,pang02,thevenet16,zhou21,hua07,wen20, arefiev15,xiao16,gong19}, the focus of our work. Among the latter, electron acceleration by resonantly excited relativistic surface plasma waves (SPW) \citep{raynaud07,riconda15,macchi18,cantono18,raynaud20,marini21,marini21b} has been demonstrated, leading to high charge, ultrashort bunches along the target surface, reaching energies largely above their quiver energy and correlated in time and space with extreme ultraviolet (XUV) harmonic emission \citep{cantono18}.  Advanced methods to control the duration and energy of the electron bunches have been proposed \citep{marini21}. As an alternative, the acceleration of electrons in the vacuum by a laser through straight energy transfer, named vacuum laser acceleration (VLA), or direct laser acceleration (DLA)~\citep{hartemann95,esarey95,pang02,hua07,wen20,arefiev15,singh22}, draws attention by its concept. New ideas to improve such a scheme have been proposed lately, such as plasma mirror injectors \citep{zhou21,thevenet16}, in which the electrons ``surf"  the reflected electromagnetic wave along a distance proportional to the Rayleigh length. The resulting bunches of nC charge reach energies of the order of MeV for a laser intensity $\sim 10^{19}$W/cm$^2$, but have significant angular divergence. Alternatively a micro-structured hollow-core target has been suggested that both guide and confine the laser pulse resulting in an enhanced and super-luminal longitudinal electric field  \citep{xiao16,gong19}.

In this Letter, a new electron acceleration mechanism is unraveled that develops when an UHI p-polarized laser pulse irradiates the wedge of an over-dense plasma target. This leads us to propose an acceleration scheme that, considering an ultra-short ($\sim 25$ fs), ultra-intense ($\sim 10^{19}\,{\rm W/cm}^2$) laser pulse (assuming micrometric wavelengths) allows to produce electron beams with 100's MeV energy, nC charge and very small (few degrees) angular aperture. The combination of high-energy, high-charge and small angular aperture makes this new scheme particularly interesting with respect to other schemes such as SPW acceleration or VLA/DLA.

\begin{figure}[b]
\begin{center}
\includegraphics[width=8.1cm]{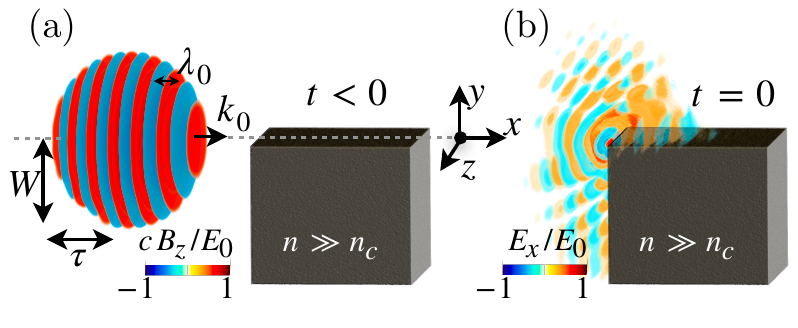}
\caption{(a) Laser-plasma interaction scheme, (b) Electric field $E_x$ extracted from the 3D PIC simulation at the time $t=0$, at which the normalized laser amplitude $a_0=5$ is maximum on the plasma edge (here $n=100n_c$).}\label{F1}
\end{center}
\end{figure}

The proposed scheme is depicted in Fig.~\ref{F1}, where the laser pulse propagates in the horizontal ($x>0$) direction. It is focused onto the wedge of the target, the latter occupying the region $x>0$ and $y<0$ and extended over several laser wavelengths in the $z$ direction. Electron acceleration occurs at the ($y=0$)$-$target surface which is irradiated by the laser at grazing incidence. In this work, we identify the key-role of the electromagnetic wave diffracted at the plasma wedge [see panel (b) in Fig.~\ref{F1}] in accelerating the electrons. For a right-angle wedge, this diffracted wave propagates cylindrically, from the wedge outwards, in all vacuum directions (from $\theta=0$ to $\theta=3\pi/2$). Most importantly, this wave carries a radial/longitudinal electric field which is responsible for the observed electron acceleration. This longitudinal field is maximum for small angles pointing in the direction of propagation of the incident laser, and it is shown to decay with the inverse square-root of the distance from the wedge. As a consequence the electron acceleration is preferably along the target surface and can be sustained over long distances. Here, sub-mm acceleration lengths are demonstrated in both 3D Particle-In Cell PIC simulations and by a simple analytical model showing that the electron energy increases with the square-root of the acceleration distance and scales linearly with the laser maximum electric field amplitude.

The effectiveness of this novel acceleration scheme is demonstrated by a 3D PIC simulation performed with {\sc Smilei}~\cite{smilei}. In this simulation, the laser pulse has a maximum normalized vector potential $a_0=eE_0/(m_e c\omega_0)=5$ ($I \lambda_0^2 =3.5\times 10^{19}\,{\rm W \mu m^2/cm^2}$, with $I$ the laser intensity and $\lambda_0$ its wavelength) a Gaussian transverse profile with waist $\sigma_0=6\,\lambda_0$, duration $\tau=8\,\lambda_0/c$ (full width at half maximum in intensity), and maximum electric field amplitude $E_0$. It is focused onto a cold plasma with electron density $n=100\,n_c$, $n_c = \epsilon_0 m_e \omega_0^2/e^2$ being the critical density beyond which the plasma is opaque to an incident laser pulse with angular frequency $\omega_0 = 2\pi c/\lambda_0$ ($\epsilon_0$ is the vacuum permittivity, $m_e$ and $-e$ the electron mass and charge, respectively, and $c$ the speed of light in vacuum). Details on the numerical parameters are given in the Supplemental Material~\cite{suppMaterial}.

Figure~\ref{F2} summarizes the 3D simulation results. Panel (a) shows in color scale the component $E_x$ of the diffracted wave (normalized to $E_0$) at time $t=18 \lambda_0/c$, $t=0$ denoting the time at which the maximum of the laser pulse reaches the edge of the target.
The electron density is also reported in gray scale. Electrons accelerated by the diffracted wave are clearly visible as bunches propagating with the longitudinal field, right above the target surface. The resulting electron energy spectrum is reported at different instants of time in panel (b), and the electron angular-energy distribution is reported at time $t = 54\lambda_0/c$ in panel (c).

\begin{figure}[t]
\begin{center}
\includegraphics[width=8.1cm]{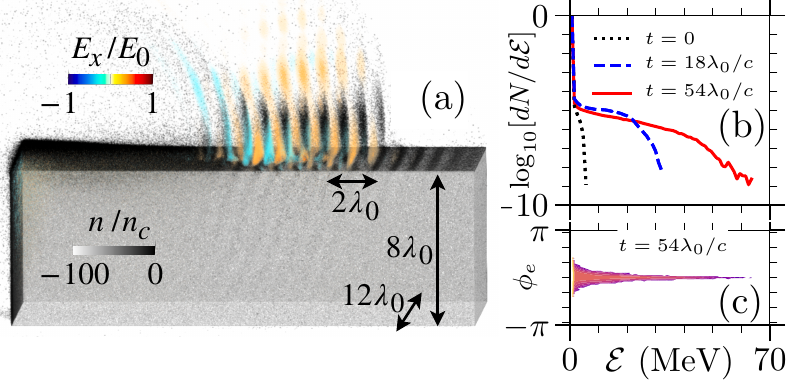}
\caption{Results from the 3D PIC simulation with $a_0=5$ and $n=100n_c$. (a) Electric field $E_x$ (in color, normalized to the maximum laser field strength $E_0$) and plasma density $n/n_c$ (in gray scale) at $t=18\lambda_0/c$. (b) Electron energy (${\cal E}$) spectrum at different times and (c) Electron energy-angular distribution at $t=54\lambda_0/c$ [$\phi_e={\rm{arctan}}(p_y/p_x)$ measures the electron direction of propagation in the ($x$,$y$) plane].} 
\label{F2}
\end{center}
\end{figure}

After only a few 10's of optical cycles of interaction, electrons have already reached energy of several tens of MeV, significantly beyond the ponderomotive energy. The charge carried by the most energetic electrons is also very large.  Considering only electrons with energy above half the maximum energy ({\it i.e.} above 30 MeV at time $t=54 \lambda_0/c$), a total electron charge of $0.8$nC is obtained (taking $\lambda_0=0.8 {\rm \mu m}$). Similar charge levels were also reported considering vacuum laser accelerators \citep{xiao16,thevenet16}, but the present scheme allows to obtain higher electron energies for similar laser intensities and duration, as well as a much narrower angular spread. Indeed, as demonstrated in panel (c), the most energetic electrons are accelerated at this time within a few degrees from the direction of the target surface.

Understanding how electrons are accelerated requires a deeper insight into the laser pulse diffraction at the plasma wedge, which can be drawn from previous theoretical~\citep{karal62,balanis} and numerical works~\citep{xi10}. In particular, different electromagnetic field components are present in the electron acceleration region ($x>0, y>0$). First, the incident electromagnetic wave. Second, considering that the plasma acts as a non-perfect conductor, a (small amplitude) SPW propagates along the vacuum-target interface. Last is the electromagnetic wave diffracted at the plasma wedge~\citep{karal62,xi10,balanis}. 

In the proposed scheme, electron acceleration is governed by the diffracted wave. Indeed, a key element for efficient electron acceleration is that, due to the non-perfectly conducting nature of the plasma, the diffracted wave carries a non-zero radial/longitudinal electric field, maximum for diffraction angle $\theta \sim 0$, pointing along the target surface and thus efficiently accelerates particles in this direction. Furthermore, unlike the SPW that is confined at the vacuum-target interface (within an evanescent length $\lambda_0$) the longitudinal field of the diffracted wave extends over few wavelengths in the ($x>0,y>0$) vacuum region (see Supplemental Material~\cite{suppMaterial}), and can thus accelerate electrons located above the target, as is shown to occur in Fig.~\ref{F2}(a).
Since the amplitude of the diffracted longitudinal field in this region decays with the inverse square-root of the distance from the wedge, it sustains the electron dynamics over long acceleration distances.

Because of the high directionality of the accelerated electrons reported in Fig.~\ref{F2}(b), we can consider the longitudinal 
electric field $E_r$ of the diffracted wave as the main driver for the electron acceleration, and approximately equal to the $E_x$ component. 
Based on these assumptions, a simple one-dimensional model can be derived to describe the electron acceleration process. We consider that the longitudinal component electric field amplitude $E_x$ decays in space as $1/\sqrt{k_0 x}$ from its maximum value $\eta E_0$ ($\eta \lesssim 1$ being the ratio of the maximum amplitude of the diffracted and the laser field, that can be extracted from the simulations), and that the wave envelope and carrier are determined by the finite laser pulse itself. As a result, the equation of motion of an electron in the resulting longitudinal field reads:
\begin{eqnarray}\label{eq:Motion}
    m_e \frac{d}{dt} \gamma v_x = -e\,\eta E_0\,e^{-(t-x/c)^2/\tau^2}\,\frac{\sin(k_0 x-\omega_0 t)}{\sqrt{k_0 x}}\,,
\end{eqnarray}
where $v_x=dx/dt$ is the electron velocity and $\gamma = (1-v_x^2/c^2)^{-1/2}$ its Lorentz factor. Equation~(\ref{eq:Motion}) can be solved numerically considering a given initial position $x_0=x(t=0)$ and zero initial velocity $v_x(t=0)=0$. It can also be solved analytically for an ultra-relativistic electron, so that $dt \sim dx/c$, and considering that the acceleration proceeds in the peak field $-\eta E_0$ ({\it i.e.} ignoring the time dependence of the field amplitude and the field oscillations), so that Eq.~(\ref{eq:Motion}) reduces to $k_0^{-1}d\gamma/dx =\eta\,a_0/\sqrt{k_0 x}$. This leads to a scaling for the electron energy
\begin{eqnarray}\label{eq:gamma_vs_t}
    \gamma(t) \sim 2\eta a_0\,\sqrt{\omega_0 t}.
\end{eqnarray}
This scaling is found to be in excellent agreement with the maximum electron energy reported in Fig.~\ref{F2}(b), which leads (for $a_0=5$, and taking $\eta=0.63$ consistent with our simulations) a maximum energy of 34~MeV ($\gamma \sim 67$) for $t=18 \lambda_0/c$ and 59~MeV ($\gamma \sim 116$) for $t=54 \lambda_0/c$. The square-root dependence of the electron energy with time is a key evidence that the acceleration takes place in the longitudinal field of the diffracted wave. 

Notice that for electrons to be accelerated by the diffracted wave, they first need to be extracted from the plasma, then injected in the wave with a (longitudinal) velocity close to $c$ so that they can phase-lock with the accelerating field. This early stage predominantly occurs at the target wedge, close to $x=0$, where its transverse electric field ($E_y$) can efficiently pull electrons out of the plasma. This happens whenever $E_y$ assumes negative values so that the electrons acquire a positive transverse velocity $v_y>0$. The resulting $v_y B_z$ contribution of the relativistic ($a_0 \gtrsim 1$) laser pulse together with the longitudinal ($E_x$) field of the diffracted wave can then bring the electron to near relativistic longitudinal velocities ($v_x \sim c$) within less than an optical cycle. This happens above a threshold in the laser intensity ($a_0 > 1$) and inspection of particle orbits shows that, while both $E_x$ and $v_y B_z$ contribute to the injection and phase locking, the first term dominates in most cases.  Moreover phase-locking requires that the electrons are generated (extracted than injected) in a region where the longitudinal electric field of the diffracted wave is negative. As reported by Karal \& Karp~\cite{karal62}, and confirmed in our simulations (see Supplemental Material~\cite{suppMaterial}), the diffracted wave is not in phase with the incident laser wave, but phase-shifted by $5\pi/4$ with respect it. As a result, injection happens only once per laser period during a time not exceeding $\lambda_0/(8c)$. This results in the nano-bunch structure of the accelerated electrons observed in Fig.~\ref{F2}(a), each bunch having a characteristic width of $\sim \lambda_0/8$ in the $x$-direction.

To test further the validity of the model developed above and the interest of the proposed acceleration scheme, we have performed a series of two-dimensional (2D) PIC simulations in the $x,y$ plane, representative of the 3D field at the center of the box. This allows  to significantly reduce the computational cost of the  simulations and  consider longer time scales and different parameters variations. In particular, the laser peak intensity was changed from $I \lambda_0^2 \simeq 3.4\times 10^{17}$ to $1.22\times10^{21}\,{\rm W \mu m^2/cm^2}$, corresponding to  $a_0$ in between $0.5$ (non-relativistic limit) and $30$, while considering otherwise unchanged laser and plasma parameters (details on the numerical parameters and consistency between 2D and 3D simulations are given the Supplemental Material~\cite{suppMaterial}).

\begin{figure}
\includegraphics[width=8.1cm]{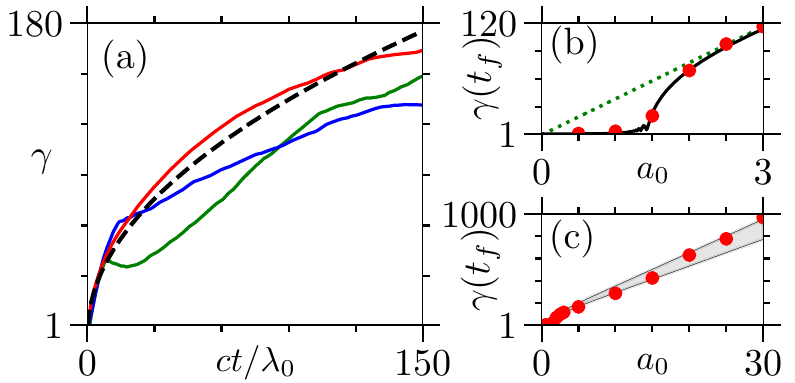}
\caption{(a) Temporal evolution of the Lorentz factor ($\gamma$) of three selected electrons extracted from 2D PIC simulations (with $a_0=5$ and $n=100n_c$); the black dashed line is obtained by solving Eq.~(\ref{eq:Motion}) numerically. 
Panels (b) and (c) report as red points the maximum Lorentz factor obtained in 2D PIC simulations at $t_f= 150\lambda_0/c$ as a function of $a_0$. In panel (a), the numerical integration of Eq.~(\ref{eq:Motion}) is shown as a black solid line, and the prediction of  Eq.~(\ref{eq:gamma_vs_t}) as a green dotted line (using $\eta=0.63$). In panel the grey region corresponds to the integration of Eq.~(\ref{eq:Motion}) for $0.6<\eta<0.75$.}\label{NF3}
\end{figure}

Figure~\ref{NF3} summarizes the results of our 2D PIC simulations. In panel (a) we show the temporal evolution of the Lorentz factor of three representative electrons (macro-particles) as they are accelerated in the diffracted wave. For this panel, $a_0=5$, and we can see that the Lorentz factor of the most energetic electron (red line) increases with time as predicted by our simple model [Eq.~(\ref{eq:Motion}) using $\eta \sim 0.63$ and $x_0=k_0^{-1}$] as shown by the dashed line. The $\sqrt{\omega_0 t}$ time-dependence is very clear for this particle. Note also that the Lorentz factor in this 2D simulation assumes the same values at times $t=18 \lambda_0/c$ ($\gamma \sim 65$) and $54 \lambda_0/c$ ($\gamma \sim 110$) than reported for the 3D simulation. The blue and green lines correspond to electrons for which phase-locking was less efficient, but that can later  be picked up by the wave and further accelerated to large energies. For all three particles, our model gives a good estimate for the maximum energy (Lorentz factor) the particle can get as a function time. This 2D simulation also shows that the acceleration can be maintained over long times, allowing to reach high energies, here of the order of $86$ MeV ($\gamma \sim 170$) at $t = 150 \lambda_0/c$ for the most energetic electron (red line). The acceleration can thus develop over long distances along the target surface. As shown in the Supplemental Material~\cite{suppMaterial}, the electrons have covered distances of the order of $150 \lambda_0$ in the $x$-direction and of a few wavelengths in the $y$-direction.

The maximum Lorentz factors achieved by an electron at time $t = 150 \lambda_0/c$ was extracted as a function of $a_0$ and reported in  Fig.~\ref{NF3}(b) and (c). Panel (b) focuses on non-relativistic to mildly relativistic field strengths, $0.5 \le a_0 \le 3$. A threshold is clearly observed, for $a_0 \sim 1.5$ (correspondingly $\eta a_0 \sim 1$). It is well reproduced by the theoretical model solving Eq.~(\ref{eq:Motion}) numerically (black solid line), while this threshold is not captured by the scaling of Eq.~(\ref{eq:gamma_vs_t}) ($\propto a_0$,  green dashed line) obtained for relativistic particles. Eq.~(\ref{eq:gamma_vs_t})  provides a quick and good  estimate for the maximum energy well above the threshold. Note also that, as $a_0$ increases, the parameter $\eta$ shows a weak dependence with $a_0\,n_c/n$, our simulations suggesting that $\eta$ increases from 0.63 up to 0.75 for $a_0 = 30$. In panel (c), the energy scaling from Eq.~(\ref{eq:Motion}) using $\eta=0.75$ is thus also reported, giving a better estimate at large $a_0$.

\begin{figure}
\includegraphics[width=8.1cm]{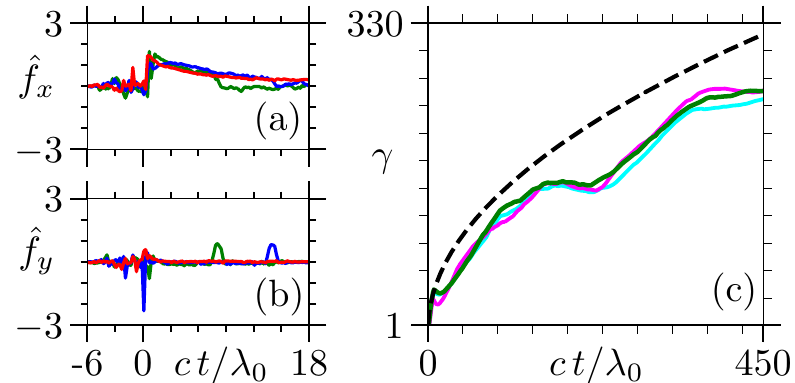}
\caption{Temporal evolution of (a) the parallel $\hat{f}_x=f_x/(m_e c\omega_0)$ and (b) the transverse $\hat{f}_y=f_y/(m_e c\omega_0)$ forces acting over the three electrons reported in Fig.~\ref{NF3}. (c) Temporal evolution of the Lorentz factor ($\gamma$) of three electrons which are accelerated over a long time interval. The black dashed line is the result from the theoretical approach using $\eta =0.63$. The green line represent the same particle in all panels. Here, $a_0=5$ and $n=100n_c$.}\label{NF4}
\end{figure}

The good agreement between the simulations and our simple one-dimensional model can be understood looking at the forces acting on the accelerated electrons. The longitudinal [$f_x = -e (E_x+v_y B_z)$] and perpendicular [$f_y = -e (E_y-v_x B_z)$] forces experienced by the three electrons discussed in Fig.~\ref{NF3} are reported in Fig.~\ref{NF4}(a) and (b), respectively. For readability, only short times $t<18 \lambda_0/c$ are shown. The longitudinal force $f_x$ in panel (a) clearly shows the $1/\sqrt{\omega_0 t}$ time-dependence expected for acceleration in the longitudinal field of the diffracted wave (with $ct \sim x$). 
This confirms the dominant contribution of $E_x$ compared to the magnetic force $v_y B_z$ (also explained as $v_y$ stays small for the high-energy electrons) and that the energy gain is provided by the work of this longitudinal field only.
From panel (b), we also get that the transverse force $f_y$ experienced by the electron is  always very small, which implies that the two contributions $E_y$ and $v_x B_z$ compensate each other ({{which is possible for $v_x \rightarrow c$}}). The transverse force assumes non negligible values only at the time of injection ($t \sim 0$) and, for the electron represented by the green and blue lines, at times $t\sim 8 \lambda_0/c$ and $15 \lambda_0/c$, respectively. A closer look at the particles orbits shows that these times correspond to the moment when those particles are bouncing off the target surface. Indeed, at those times the electrons penetrate the plasma skin depth, experience a screened electric field ($E_y \rightarrow 0$) and are turned back by the strong $v_x B_z$ force ($B_z$ is not screened). At those times, the electrons do not gain energy [see Fig.~\ref{NF3}(a)], but they can reenter the wave and get further accelerated.

The slow decrease ($\propto 1/\sqrt{k_0 x}$) of the longitudinal field of the diffracted wave along the target surface means that electron can, in principle, remain in phase and be accelerated over distances/times even longer that what considered so far. This is confirmed in Fig.~\ref{NF4}(c) where few selected high-energy electrons from our reference case were tracked over 450 optical cycles, propagating distances $\sim 450 \lambda_0$  along the target surface and reaching energies of nearly 130 MeV ($\gamma \sim 260$). 
To understood this we can examine the phase-shift an electron acquires with respect to the accelerating wave 
\begin{eqnarray}
\Delta\varphi = k_0 \int_{t_{\rm inj}}^t\,(c-v_x)\,dt\,.
\end{eqnarray}
Considering relativistic electrons only, $v_x(t)$ is well approximated by Eq.~(\ref{eq:gamma_vs_t}) [using $ v_x=(1-1/\gamma^2)^{1/2}$]. 
Assuming that the electron energy at time $t$ is much larger than that at the moment of injection [$\gamma(t) \gg \gamma(t_{\rm inj})$], one obtains that the phase-shift $\Delta\varphi \sim (8\eta^2a_0^2)^{-1}\ln 4\eta^2 a_0^2\,\omega_0 t$ increases logarithmically with time. Conversely, the dephasing time $\omega_0 t_d \propto \exp(8\pi \eta^2 a_0^2)$, at which $\Delta\varphi \sim \pi$, increases exponentially with $\eta^2 a_0^2$.
This ensures that the electron can remain in phase with the accelerating field whenever $\eta a_0>1$.

Considering the forces in Fig \ref{NF4}(a) and (b), the transverse motion in the complex field resulting from the superposition of the cylindrical diffracted wave and incident laser field can induce in some particles a temporary dephasing with respect to the accelerating field as they hit the surface. A similar effect can occur after  much longer time scales, and it is at the origin of the fact that for a short amount of time the energy stops increasing, as visible for example in  Fig.~\ref{NF4}(c), but starts again after re-injection.
Note also that 3D diffraction effects can set in over long distances, and limit the acceleration.

The robustness of the proposed acceleration mechanism was tested using complementary 2D PIC simulations (not shown). In these simulations, various parameters were changed, such as the shape of the target's corner or the roughness of the plasma surface. The presence of a small pre-plasma, or a small misalignment of the laser were also considered (the laser impinging at an angle up to $\pm 5^{\circ}$ with respect to the surface, and with variations of the focus location of a few wavelengths in all directions).
In all these complementary simulations, electron acceleration was shown to be marginally impacted, suggesting that this acceleration scheme could be easily implemented in experiments aimed at demonstrating new energetic particles sources. A consequence of the robustness of this acceleration mechanism is that, even though it was never identified or discussed in previous works, acceleration in the field of the diffracted could develop in various laser-plasma interaction setups~\cite{noteShen}.

In conclusion, a new mechanism of electron acceleration has been identified in the interaction of a relativistically intense laser pulse with an overdense plasma wedge. Both 3D and 2D PIC simulations have shown this mechanism to be robust and provide highly charged (nC), well collimated  electron bunches with energies of several tens to hundreds of MeV.
A simple analytical model has been developed that shows that the maximum energy of the accelerated electrons scales linearly with the laser field strength parameter ($a_0$) and increases with the square-root of time to values well beyond the ponderomotive scaling.  From this model, we obtain that the particles energy gain can be controlled by the longitudinal target size, and in particular that the maximum electron energy scales with the square-root of this size. As for the total charge of the accelerated beam, it can be controlled by the laser pulse duration: the longer the pulse being, the more electron nano-bunches being accelerated.

The simplicity and robustness of the proposed accelerating scheme pave the way to new experiments that can be easily done on current laser facilities.

\section*{Acknowledgement}
P.S.K. was supported by the CEA NUMERICS program, which has received funding from the European Union's Horizon 2020 research and innovation program under the Marie Sklodowska-Curie grant agreement No. 800945. Financial support from Grant No. ANR-11-IDEX-0004-02 Plas@Par is acknowledged. Simulations were performed on the Irene-SKL machine hosted at TGCC- France, using High Performance Computing resources from GENCI-TGCC (Grant No. 2021-x2016057678).
The authors are grateful to F. Amiranoff, A. Grassi, F. Massimo, A. Mercuri, F. P\'erez, L. Romagnani, and T. Vinci for fruitful discussions and to the SMILEI dev-team for technical support.

\end{document}


\title{Supplemental Material: Electron acceleration by laser plasma wedge interaction}

\date{\today}

\author{S. Marini} 
\affiliation{LSI, CEA/DRF/IRAMIS, CNRS, \'Ecole Polytechnique, Institut Polytechnique de Paris, F-91128 Palaiseau, France.}
\affiliation{LULI, Sorbonne Universit\'e, CNRS, CEA, \'Ecole Polytechnique, Institut Polytechnique de Paris, F-75252 Paris, France.}

\author{P. S. Kleij} 
\affiliation{LSI, CEA/DRF/IRAMIS, CNRS, \'Ecole Polytechnique, Institut Polytechnique de Paris, F-91128 Palaiseau, France.}

\author{M. Grech} 
\affiliation{LULI, CNRS, CEA, Sorbonne Universit\'e, \'Ecole Polytechnique, Institut Polytechnique de Paris, F-91120 Palaiseau, France.}

\author{M. Raynaud} 
\affiliation{LSI, CEA/DRF/IRAMIS, CNRS, \'Ecole Polytechnique, Institut Polytechnique de Paris, F-91128 Palaiseau, France.}

\author{C. Riconda \footnote{caterina.riconda@upmc.fr}}
\affiliation{LULI, Sorbonne Universit\'e, CNRS, CEA, \'Ecole Polytechnique, Institut Polytechnique de Paris, F-75252 Paris, France.}

\begin{abstract}
\end{abstract}

\maketitle

\section{I.~Numerical set-up}

Simulations have been performed with the SMILEI particle-in-cell (PIC) code~\citep{smilei}. The 3D simulation box is $72\lambda_0 \times 24\lambda_0 \times 24\lambda_0$ (in the $x$-$y$-$z$ directions) and composed of $4608 \times 1536 \times 1536$ cells (spatial resolution $\Delta=\lambda_0/64$). The time  resolution is $\Delta t = 0.95\Delta/(\sqrt{3}c)$. Electromagnetic field boundary conditions are injecting/absorbing in the $x$-direction and absorbing in the $y,~z$-directions. Particle boundary conditions are either thermalizing at $y_{\rm min}$, $z_{\rm min}$, and $x_{\rm max}$ or absorbing in the complementary directions. In each computational plasma cell, there are $4$ macro-electrons and $4$ macro-ions. The ion over electron mass ratio is given by  $A~m_p/Z=1836 ~m_e$, with $A,Z$ respectively the atomic number and charge, and $m_p$ the proton mass.

The 2D simulation box is $192\lambda_0 \times 36\lambda_0$ (in the $x,y$- directions) and composed of $49152 \times 9216$ cells (spatial resolution $\Delta=\lambda_0/128$). The time resolution is $\Delta t = 0.95\Delta/(\sqrt{2}c)$. In each computational plasma cell, there are $32$ macro-electrons and $32$ macro-ions. The electromagnetic field and particle boundary conditions are the same as in the 3D simulation.

\section{II. Electromagnetic fields}

\begin{figure}[b]
\begin{center}
\includegraphics[width=4.cm]{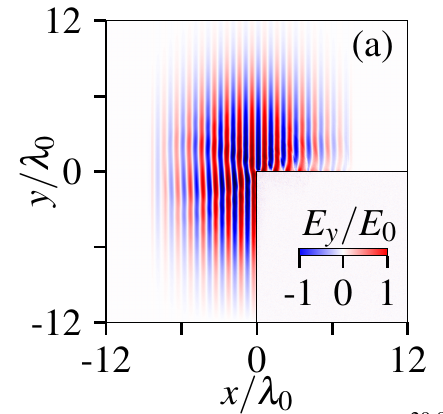}
\includegraphics[width=4.cm]{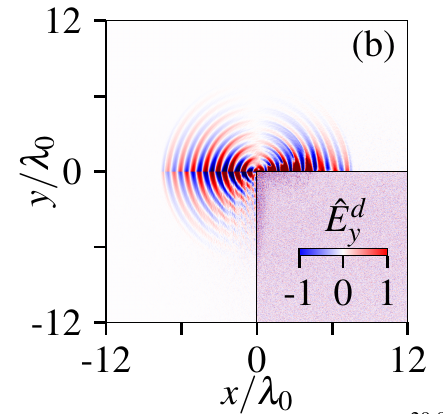}
\includegraphics[width=4.cm]{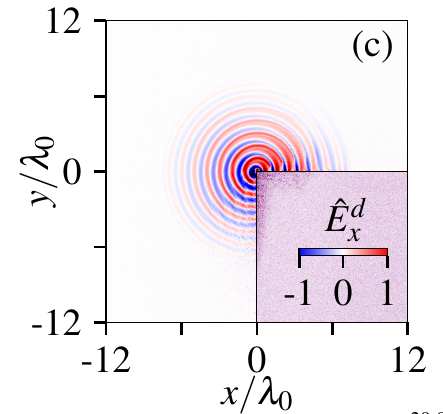}
\includegraphics[width=4.cm]{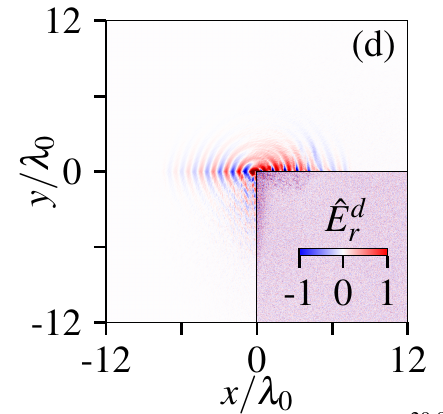}
\caption{Fields extracted from the 3D PIC simulation at $t=0$: 
(a) total electric field $E_y/E_0$, (b) and (c) $y$- and $x$-components of the electric field associated to the diffracted wave, respectively, and (d) radial component of the electric field $\hat{E}_r^d$ associated to the diffracted wave.
}
\label{S1}
\end{center}
\end{figure}

Figure \ref{S1} reports on 3D simulation results for $a_0=5$ and $n=100n_c$ at time $t=0$ (when the laser field amplitude on target is maximum) and for $z=0$ (center of the laser pulse in the third dimension). It shows (a) the total electric field $E_y/E_0$, (b) the $E_y$ and (c) $E_x$ components of the diffracted wave only, and (c) the radial component $E_r$ of the diffracted wave only:
\begin{center}
$\hat{E}_r^d=\hat{E}_x^d (x/\sqrt{x^2+y^2})+\hat{E}_y^d (y/\sqrt{x^2+y^2})$\,
\end{center}
all electric fields shown with a hat are reported in units of $m_e c \omega_0/e$. 

The field structures observed here are similar to those reported in the literature~\cite{karal62,xi10,balanis} considering the irradiation of a right-angled wedge by a plane wave. Our 3D PIC simulations confirm that even at relativistic intensities, and considering finite size and pulse duration, the diffracted field properties are similar to those described in the literature.\\

\begin{figure}[t]
\begin{center}
\includegraphics[width=4.cm]{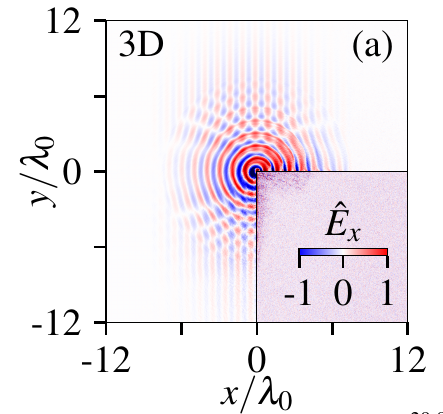}
\includegraphics[width=4.cm]{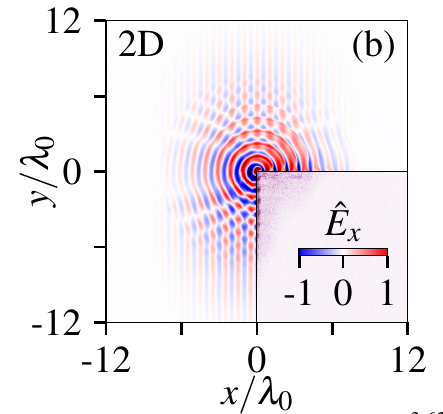}
\caption{Comparison of the total field $\hat{E}_x$ extracted from (a) 3D and (b) 2D simulations.}
\label{S2}
\end{center}
\end{figure}

\begin{figure}[ht]
\begin{center}
\includegraphics[width=4.cm]{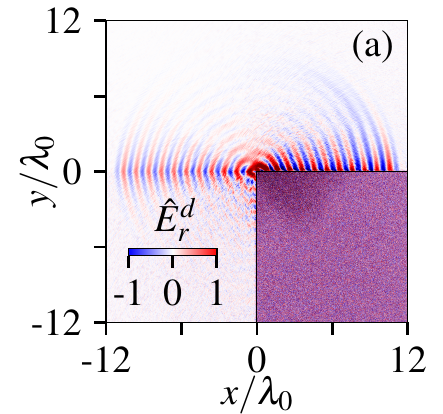}
\includegraphics[width=4.cm]{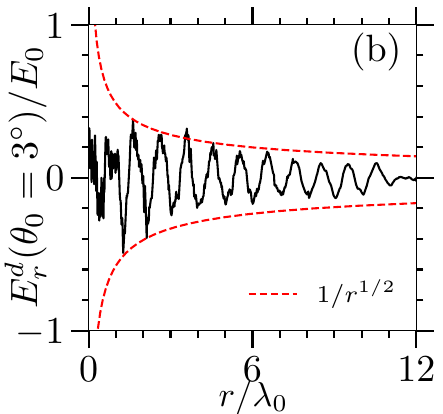}
\caption{2D PIC simulation using a constant laser temporal profile. 
(a) Radial field $\hat{E}_r^d$ associated to the diffracted wave. (b) Line-out of $\hat{E}_r^d$ recorded at an angle $\theta_0=3^\circ$ from the target surface. 
}
\label{S3}
\end{center}
\end{figure}

The same electromagnetic field structures are recovered in 2D simulations using the same physical (laser and plasma) parameters, but with higher resolution, box size and duration of the simulation. A direct comparison of the $\hat{E}_x$ component of the electric field drawn from 3D and 2D PIC simulations is reported in Fig.~\ref{S3}. The two simulations show excellent agreement, and we verified that the same is true for the other fields.

Note also that a 3D simulation with reduced resolution (not shown) confirmed that both the general electromagnetic field structure and electron acceleration were correctly described in 2D up to the maximum time accessible in 3D $t \sim 150\lambda_0/c$.\\

Last, Fig. \ref{S3} reports on the result of a 2D simulation for which the laser field amplitude was kept constant over the whole simulation duration. This allows us to highlight the decrease with the distance from the wedge of the longitudinal field $E_r$ close to the surface.
The longitudinal field of the diffracted wave $\hat{E}_r^{d}$ is reported in panel (a), while panel (b) shows a line-out of the field recorded at a small angle $\theta_0 = 3^\circ$ with respect to $x$-direction (to remove noisy contributions at the surface location). The one-over-square-root dependence expected from \citep{karal62,xi10,balanis} is recovered.

\section{III. Fields and particle acceleration}

\begin{figure}[t]
\begin{center}
\includegraphics[width=15cm]{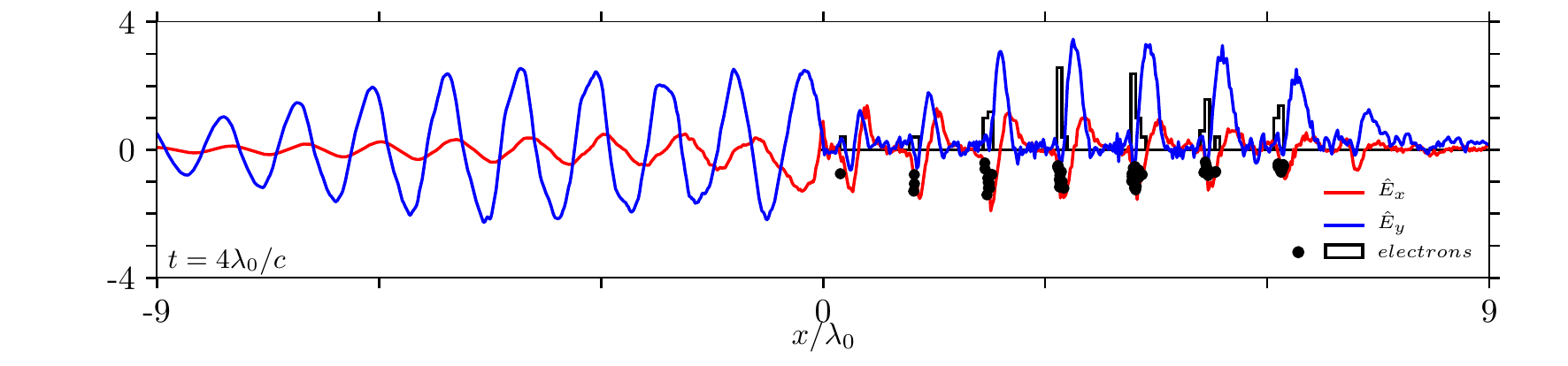}
\caption{Total transverse (blue) and longitudinal (red) electric fields at time $t=4\lambda_0/c$ as extracted from a 2D simulation with $a_0=5$ and $n=100 n_c$. Black dots represent high energetic test electrons propagating with the wave and the solid black line is a histogram representing the number of accelerated electrons at a given position.}
\label{S4}
\end{center}
\end{figure}

Figure~\ref{S4} shows the (total) longitudinal and transverse electric fields measured at the plasma surface, $y/\lambda_0=0$. Electrons are injected at the plasma wedge, $x=0$, when both $\hat{E}_x$ (solid red line) and $\hat{E}_y$ (solid blue line) are negative. Because of a phase-shift of $5\pi/4$ between the $E_y$ and $E_r$ components for $x>0$, 
electrons are injected once per laser period as nano-bunches accelerating in the region of negative $\hat{E}_x$ field. In this Figure, the black dots represent high energetic test electrons propagating with the electromagnetic wave. The solid black line is a histogram representing the number of electrons at a given position. It shows that electrons form nano-bunches with the typical size $\sim \lambda_0/8$ consistent the $5\pi/4$ phase-shift between the $E_y$ and $E_x$ electric fields.\\

\begin{figure}[t]
\begin{center}
\includegraphics[width=8.cm]{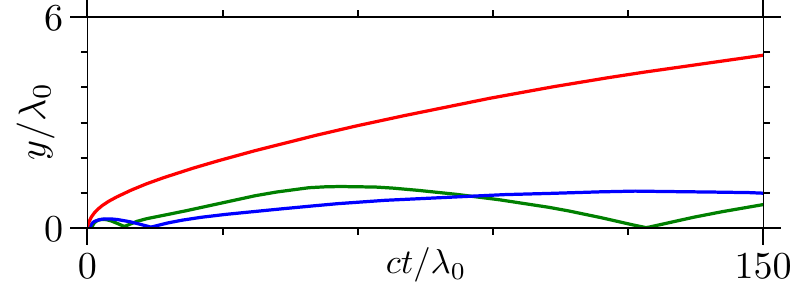}
\caption{Longitudinal excursions of the electrons shown in Fig. 3. of the manuscript.}
\label{S5}
\end{center}
\end{figure}

Finally, Fig.~\ref{S5} shows the transverse excursions of the 3 electrons presented in Fig. 3 of the manuscript. While these electron trajectories are followed over a long time $150 \lambda_0/c$ and electrons have thus propagated over $\sim 150 \lambda_0$, their transverse excursion is limited to a few wavelengths at most. This confirms that, once the particle has phase-locked with the accelerating wave, its transverse displacement is negligible with respect to the longitudinal one ($x \sim ct$). Note however that, in particular for the red curve which represents the most energetic electron (it reached energies $\sim 85$ MeV at $t=150\lambda_0/c$), the particle is still gaining energy while at several wavelengths above the target surface. This excludes surface plasma waves as the origin of the particle acceleration as these waves are well localized at the target surface and decay exponentially with the distance from the surface over evanescent lengths of the order of $\lambda_0$. To further confirm this statement, we performed complementary simulations (not shown here) with different plasma densities. These simulations showed the same behavior for the electron acceleration regardless of the density values. In contrast, electron energy gain by SPW depends on the wave phase velocity and is strongly tied to the plasma density \citep{riconda15}.

